\documentclass[10pt]{article}
\usepackage[T1]{fontenc}
\usepackage[latin9]{inputenc}
\usepackage{geometry}
\usepackage{amsmath}
\usepackage{graphicx}
\usepackage{setspace}
\usepackage{amssymb}
\usepackage{esint}

\makeatletter

\usepackage[T1]{fontenc}
\usepackage{ae}
\usepackage{aecompl}

\usepackage[md]{titlesec}

\usepackage[hang,footnotesize]{caption} 

\providecommand{\tr}{\mathrm{tr} }

\AtBeginDocument{
  
}

\makeatother

\begin{document}
\begin{flushright}
BROWN-HET 1493
\par\end{flushright}

~

\begin{onehalfspace}
\begin{center}
{\huge Six-Dimensional Yang Black Holes }\\
{\huge in Dilaton Gravity}
\par\end{center}{\huge \par}
\end{onehalfspace}

~

\begin{onehalfspace}
\begin{center}
Michael C. Abbott and David A. Lowe\\
\emph{Brown University, Providence RI, USA.}\\
\emph{abbott, lowe@brown.edu}
\par\end{center}
\end{onehalfspace}

~

\begin{quote}
We study the six-dimensional dilaton gravity Yang black holes of \cite{Bergshoeff:2006bs},
which carry $(1,-1)$ charge in $SU(2)\times SU(2)$ gauge group.
We find what values of the asymptotic parameters (mass and scalar
charge) lead to a regular horizon, and show that there are no regular
solutions with an extremal horizon. 
\end{quote}
~

\section{Introduction}

The canonical charged black hole is the Reissner--Nordstr\"om (RN)
solution. For a given charge and mass, the no-hair theorem guarantees
it is the only static solution of Einstein--Maxwell theory in four
spacetime dimensions. It can carry magnetic monopole charge, with
its singularity hidden behind a horizon, which non-gravitating objects
cannot do without a naked singularity. It has both an inner and outer
horizon (an event horizon, and an inner Cauchy horizon) which merge
in the extremal limit $Q=M$, leading to an $AdS$ throat geometry. 

It is interesting to consider the generalization to nonabelian gauge
groups. Without gravity, theories with nonabelian gauge groups have
Yang monopoles, which like their Dirac cousins are singular \cite{Dirac:1931kp,Yang:1977qv}.
But when the gauge group is spontaneously broken, there are non-singular
't~Hooft--Polyakov monopoles \cite{tHooft:1974qc,Belavin:1975fg}.
Adding gravity, one can place a small Schwarzschild black hole at
the centre of one of these, to produce one of Weinberg's hairy black
holes \cite{Lee:1991jw,Lee:1991vy,Weinberg:2001gc}. These can have
identical mass and charge to the RN solutions, which still exist in
the larger theory, hence forming classical hair. Gravity also allows
other regular solutions which do not exist in flat space \cite{Bartnik:1988am}.

String theory encourages us to examine theories in more than four
dimensions, possibly with a dilaton. Extra dimensions allow black
holes to have novel features, such as multiple angular momenta \cite{Myers:1986un}
and non-spherical horizon topology \cite{Emparan:2001wn,Elvang:2007rd}.
Adding charge, \cite{Gibbons:2006wd} study $SO(2k+1)$ gauge theory
in $2k+2$ dimensions, with gravity. This includes RN as the $k=1$
case, and the $k=2$ case also has a double horizon and an extremal
limit. See also \cite{Zeng:2006ej,Tchrakian:2006nn}. 

In addition to these string-inspired solutions, there are some which
arise more directly in string theory. Fundamental heterotic strings
can end on certain monopoles, \cite{Polchinski:2005bg,Bergman:2006qs}
and various kinds of D-branes can end on others \cite{Strominger:1995ac,Townsend:1995af,Berman:2007bv}.
In this paper, we study the case of Bergshoeff, Gibbons \& Townsend
\cite{Bergshoeff:2006bs} in which an M5-brane connects two M9-branes
which are separated along the 11th direction. Taking the 10-dimensional
view, there are 4 dimensions along the intersection, 6 perpendicular
to it, and a dilaton. The two M5-M9 intersections give opposite charges
in two copies of $SU(2)$, which are superimposed. The same monopole
can also be constructed using instead the D-branes of type IIA string
theory \cite{Belhaj:2007yw}.

Section \ref{sec:Asymptotic-Behaviour} sets up the problem, and recalls
the asymptotic expansion given by \cite{Bergshoeff:2006bs}. In Section
\ref{sec:Near-the-Regular} we find a near-horizon expansion, and
connect this to a subspace of the parameter space at infinity. Section
\ref{sec:Closed-Geometry,-a} examines a singular case, and finally
Section \ref{sec:Near-the-singularity} studies the solution near
the singularity, connecting this to the near-horizon solution as a
check that nothing unexpected happens in between.

\section{Asymptotic Behaviour\label{sec:Asymptotic-Behaviour}}

The action for 6-dimensional dilaton gravity with a Yang-Mills field
is \cite{Bergshoeff:2006bs}\[
S=\int d^{6}x\sqrt{-\det g}\left[R-(\partial\sigma)^{2}-4\kappa e^{-\sigma}\tr\left|F^{2}\right|\right]\]
and we immediately choose units such that $\kappa=1$. Write the metric
as \begin{equation}
ds^{2}=-e^{2\lambda}\Delta dt^{2}+\frac{dr^{2}}{\Delta}+r^{2}d\Omega^{2}\label{eq:metric-ansatz}\end{equation}
in terms of two functions of the radial co-ordinate, $\lambda(r)$
and $\Delta=1-2\mu(r)/r^{3}$. 

We take the  gauge group to be $SU(2)\times SU(2)$. Yang monopoles
may then have charges in one or both factors. For monopole with charge
$(1,-1)$, $\tr\left|F^{2}\right|=6/r^{4}$ and the equations of motion
become \cite{Bergshoeff:2006bs}\begin{align}
0 & =\Delta r^{4}\sigma''+\left(4r^{3}-2\mu-6e^{-\sigma}r\right)\sigma'+12e^{-\sigma}\nonumber \\
0 & =\mu'-3e^{-\sigma}-\Delta r^{4}(\sigma')^{2}/8\label{eq:eleven+twelve+thirteen}\\
0 & =4\lambda'-(\sigma')^{2}r\,.\nonumber \end{align}
The third equation simply fixes $\lambda$ in terms of $\sigma$,
up to one constant, which is an overall scaling of $t.$ Whenever
an aymptotic region exists, we fix this by demanding $\lambda(\infty)=0$.
The meat is in the first two equations, which are second order in
$\sigma$ and first in $\mu$, thus we need 3 other boundary conditions. 

Reference \cite{Bergshoeff:2006bs} gives the following asymptotic
solution for large $r$, with three parameters $\sigma_{0}$, $\mu_{0}$
and $\Sigma$:\begin{align}
\sigma & =\sigma_{0}+\frac{A}{r^{2}}+\frac{\Sigma}{r^{3}}+\frac{A^{2}}{2r^{4}}+\ldots\nonumber \\
\mu & =\frac{Ar}{2}+\mu_{0}-\frac{A\Sigma}{2r^{2}}-\frac{4A^{3}+9\Sigma^{2}}{24r^{3}}+\ldots\label{eq:asymp-sol}\\
\lambda & =-\frac{A^{2}}{r^{4}}-\frac{3A\Sigma}{5r^{5}}+\ldots\qquad\qquad\qquad\mbox{with }A=6e^{-\sigma_{0}}\,.\nonumber \end{align}

Notice in the above equations of motion that the existence of a regular
horizon ($\Delta(r_{H})=0$ with $\Delta r^{4}\sigma''=0$) imposes
one relation between $\sigma'$ and $\mu$ at $r_{H}$, thus reducing
the number of free parameters by one. Thus one would expect only a
two-parameter family of these asymptotic solutions to lead to a regular
horizon. In the next section we find these solutions.

\section{Near the Regular Horizon\label{sec:Near-the-Regular}}

\noindent To find solutions with a regular horizon, we assume its
existence at $r=r_{H}$ and expand in $d=r-r_{H}$. This leads to
the following near-horizon solution, with two parameters $r_{H}$
and $\sigma_{H}$ (plus a third, $\lambda_{H}$, which is fixed by
$\lambda(\infty)=0$):\begin{align}
\mu & =\frac{r_{H}^{3}}{2}+\frac{A_{H}}{2}d+\frac{3A_{H}^{2}}{4r_{H}(3r_{H}^{2}-A_{H})}d^{2}+\ldots\nonumber \\
\sigma & =\sigma_{H}-\frac{2A_{H}}{r_{H}(3r_{H}^{2}-A_{H})}d+\frac{A^{3}-12A_{H}^{2}r_{H}^{2}+18Ar_{H}^{4}}{r_{H}^{2}(3r_{H}^{2}-A_{H})^{3}}d^{2}+\ldots\label{eq:NH-sol}\\
\lambda & =\lambda_{H}+\frac{A_{H}^{2}}{r_{H}(3r_{H}^{2}-A_{H})^{2}}d+\ldots\qquad\qquad\qquad\mbox{with }A_{H}=6e^{-\sigma_{H}}\,.\nonumber \end{align}
\begin{figure}
\includegraphics[width=0.5\columnwidth]{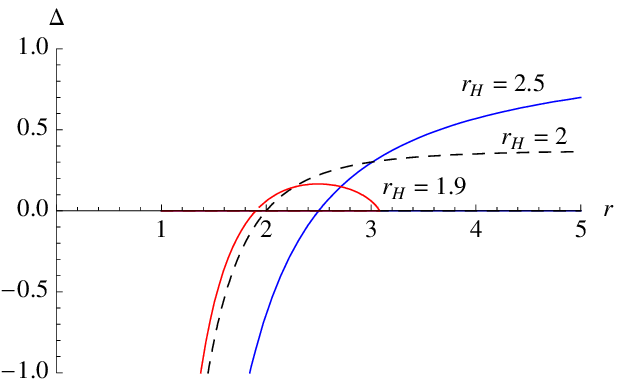}\includegraphics[width=0.5\columnwidth]{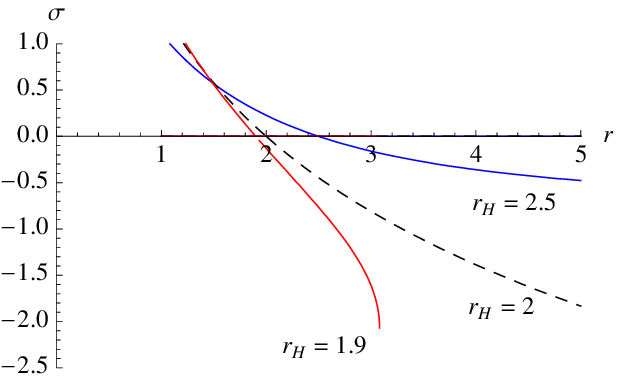}\caption{Plots of $\Delta(r)$ and $\sigma(r)$ for three values of $r_{H}$,
all with $\sigma_{H}=0$. They are $r_{H}=1.9$ (red), $r_{H}=2$
(dashed) and $r_{H}=2.5$ (blue, extending to $r=\infty$).\label{fig:plots-delta-sigma}}

\end{figure}

Given one solution, we can generate another by scaling by a factor
$\gamma$: the new functions defined by \begin{align}
\bar{\mu}(\bar{r}) & =\frac{\mu(\gamma\bar{r})}{\gamma^{3}}\qquad\mbox{i.e. }\bar{\Delta}(\bar{r})=\Delta(\gamma\bar{r})\nonumber \\
\bar{\sigma}(\bar{r}) & =\sigma(\gamma\bar{r})+2\log\gamma\label{eq:scaling-func}\\
\bar{\lambda}(\bar{r}) & =\lambda(\gamma\bar{r})\nonumber \end{align}
obey the same equations.%
\footnote{Alternatively, we could write this scaling as a set of replacements\begin{align}
r & \to\gamma\bar{r}\qquad\qquad\mbox{thus }\frac{d}{dr}\to\gamma^{-1}\frac{d}{d\bar{r}}\nonumber \\
\sigma & \to\bar{\sigma}-2\log\gamma\label{eq:scaling-arrows}\\
\mu & \to\gamma^{3}\bar{\mu}\qquad\qquad\mbox{i.e. }\Delta\to\bar{\Delta}\nonumber \\
\lambda & \to\bar{\lambda}\nonumber \end{align}
which leave equations \eqref{eq:eleven+twelve+thirteen} unchanged. %
} Starting with the near-horizon solution $(r_{H},\sigma_{H})$, the
new solution will have parameters \begin{equation}
(\bar{r}_{H},\bar{\sigma}_{H})=(r_{H}/\gamma\,,\:\sigma_{H}+2\log\gamma).\label{eq:scaling-NHparam}\end{equation}
We can set $\bar{\sigma}_{H}=0$ by taking $\gamma=e^{-\sigma_{H}/2}$.
Thus we may focus on the case $\sigma_{H}=0$ ($A_{H}=6$) without
loss of generality. 

The behavior of this solution is different in three ranges of $r_{H}$:

\begin{itemize}
\item $r_{H}<\sqrt{2}$ implies $\Delta'<0$ at the horizon, so this cannot
be the outermost horizon (as $\Delta=1$ at infinity). 
\item $\sqrt{2}<r_{H}<2$ leads to a singular space: at some finite $r=r_{M}$,
$\Delta\to0$ and $\sigma'\to-\infty$, producing a curvature singularity.
This is discussed in the next section.
\item $r_{H}>2$ matches onto the asymptotic solution \eqref{eq:asymp-sol},
with $\Delta\to1$ as $r\to\infty$.
\end{itemize}
Figure \ref{fig:plots-delta-sigma} shows the second and third cases,
and the boundary between them. 

\begin{figure}
\includegraphics[width=0.5\columnwidth]{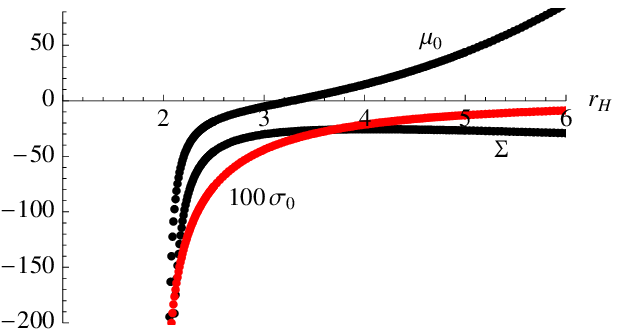}\includegraphics[width=0.5\columnwidth]{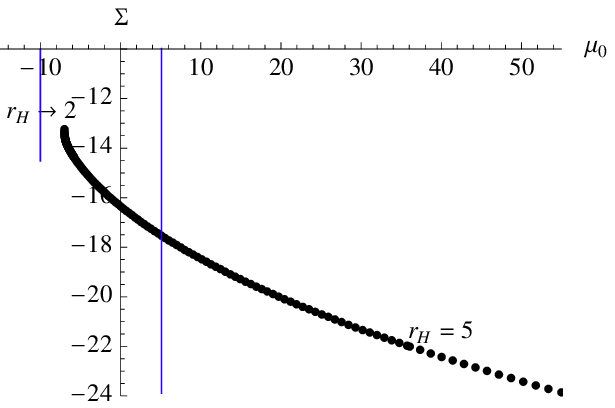}\caption{Parameters of the asymptotic solution \eqref{eq:asymp-sol} connecting
to near-horizon solution \eqref{eq:NH-sol} with $\sigma_{H}=0$.
On the left, $\mu_{0}$, $\Sigma$ and $100\sigma_{0}$ (in red) against
$r_{H}$. On the right, the scaling \eqref{eq:scaling-Aparam} has
been used to set $\sigma_{0}$ to zero, and $\Sigma$ is plotted against
$\mu_{0}$. The blue lines are regions investigated in section \ref{sec:Near-the-singularity}
below. \label{fig:parameters-at-infinity}}

\end{figure}

We now study the third case. Since an asymptotic region exists, we
can fix $\lambda_{H}$ using $\lambda(\infty)=0$. The remaining near-horizon
parameters $(r_{H},\sigma_{H})$ are mapped to a two dimensional subspace
of $(\mu_{0},\Sigma,\sigma_{0})$ at infinity. Focusing on the $\sigma_{H}=0$
case,  $r_{H}$ is mapped to a line. The first part of Figure \ref{fig:parameters-at-infinity}
shows these parameters along this line.

The scaling \eqref{eq:scaling-arrows}, acting on the asymptotic solution's
parameters, reads \begin{equation}
(\bar{\mu}_{0},\bar{\Sigma},\bar{\sigma}_{0})=(\gamma^{-3}\mu_{0}\,,\:\gamma^{-3}\Sigma,\:\sigma_{0}+2\log\gamma).\label{eq:scaling-Aparam}\end{equation}
Just as we could at the horizon, we can set $\bar{\sigma}_{0}=0$
by taking $\gamma=e^{-\sigma_{0}/2}$. Then the line of parameters
connecting to a regular horizon $r_{H}>2$ becomes a line in the $(\bar{\mu}_{0},\bar{\Sigma})$
plane.  The second part of Figure \ref{fig:parameters-at-infinity}
shows this curve. Asymptotic solutions with (scaled) parameters not
on this curve have a naked singularity at the origin.

\section{Closed Geometry, a singular case\label{sec:Closed-Geometry,-a}}

Here we study the second case above, $\sqrt{2}<r_{H}<2$, in which
$\Delta$ appears to have a second zero outside the regular horizon. 

Begin by changing co-ordinates from $r$ to radial distance $y$:\begin{equation}
ds^{2}=-a(y)\, dt^{2}+dy^{2}+r(y)^{2}d\Omega^{2}\label{eq:new-metric}\end{equation}
replacing $\lambda(r)$, $\mu(r)$ and $\sigma(r)$ with new functions
$a(y)$, $r(y)$ and $\sigma(y)$. Among other relationships between
them,\[
\Delta(r)=\left(\frac{dr}{dy}\right)^{2}\qquad\mbox{ and }\qquad\sigma'(r)=\frac{d\sigma}{dy}\bigg/\frac{dr}{dy}.\]
Thus a maximum of $r(x)$, with $\sigma(x)$ regular, will lead to
$\Delta=0$ and $\sigma'(r)$ diverging, which is what happens in
Figure \ref{fig:plots-delta-sigma}. 

Placing $y=0$ at this point, we expand and find the following series
solution with parameters $r_{M}$, $\sigma_{M}$ and $\sigma_{2}$:
\begin{align}
r & =r_{M}-\frac{3(2A_{M}-r_{M}^{2})}{2r_{M}^{2}}y^{2}+\ldots\qquad\qquad\mbox{with }A_{M}=6e^{-\sigma_{M}}\nonumber \\
\sigma & =\sigma_{M}+\frac{\sqrt{12}\sqrt{2A_{M}-r_{M}^{2}}}{r_{M}^{2}}y+\frac{\sigma_{2}}{2}y^{2}+\ldots\label{eq:M-sol}\\
a & =4\log\left|y\right|+a_{M}-\frac{2r_{M}\sigma_{2}}{\sqrt{3}\sqrt{2A_{M}-r_{M}^{2}}}y+\ldots\,.\nonumber \end{align}
This matches the numerical solution very well. So as far as the first
two of equations \eqref{eq:eleven+twelve+thirteen} are concerned,
the apparent divergences at $r_{M}$ are simply due to this being
a maximum radius for the transverse sphere, beyond which $r$ is no
longer a valid co-ordinate.

But the third equation (for $\lambda$, or $a$) then leads us to
a different interpretation: since $a=-g_{00}$ diverges, the point
$r_{M}$ is in fact a curvature singularity. Hence we discard these
solutions.

\section{Near the Singularity\label{sec:Near-the-singularity}}

We claimed above that for generic values of the asymptotic parameters
(i.e. those not on the curve of Figure \ref{fig:parameters-at-infinity})
there is a naked singularity, and that there is never a second horizon
inside the regular horizon studied. 

Here we justify these claims by connecting the near-horizon and asymptotic
solutions to another approximate solution near the singularity at
$r=0$. This is the following: \begin{align}
\sigma & =\alpha\log r+\sigma_{Z}+\ldots\nonumber \\
\mu & =\frac{\beta}{r^{\alpha^{2}/4}}+\frac{12e^{-\sigma_{Z}}\: r^{1-\alpha}}{(\alpha-2)^{2}}+\frac{\alpha^{2}\: r^{3}}{2(\alpha^{2}+12)}+\ldots\label{eq:Z-sol}\\
\lambda & =\frac{\alpha^{2}}{4}\log r+\lambda_{Z}+\ldots\,.\nonumber \end{align}
(We discuss further corrections to this in the Appendix.)

\begin{figure}
\includegraphics[width=0.5\columnwidth]{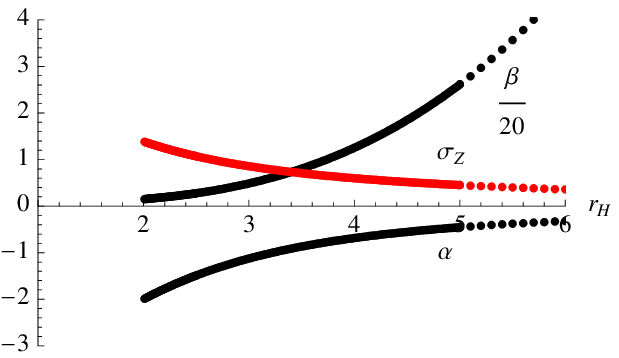}\includegraphics[width=0.5\columnwidth]{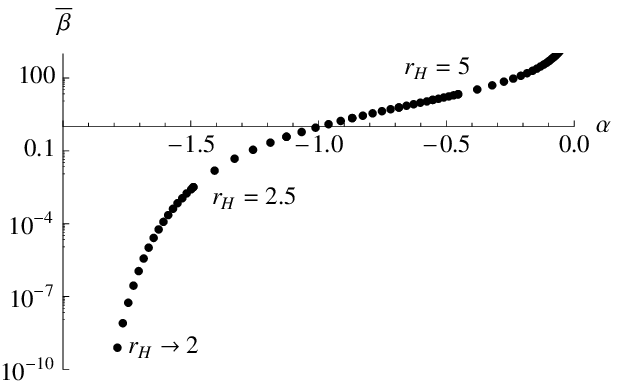}\caption{Parameters of the solution near the singularity \eqref{eq:Z-sol}
connecting to near-horizon solution \eqref{eq:NH-sol} with $\sigma_{H}=0$.
On the left, $\alpha$, $\beta/20$ and $\sigma_{Z}$ (in red). On
the right, the scaling \eqref{eq:scaling-Zparam} has been used to
set $\sigma_{Z}$ to zero, and we now plot $\overline{\beta}$ against
$\alpha$ for $2<r_{H}<20$.\label{fig:parameters-at-zero}}

\end{figure}

Just as we integrated outwards to connect the near-horizon solution
to the asymptotic solution above, we can integrate inwards and connect
it to the solution near the singularity. Figure \ref{fig:parameters-at-zero}
is analogous to Figure \ref{fig:parameters-at-infinity}.

Acting on this solution's parameters, the scaling \eqref{eq:scaling-arrows}
now reads

\begin{equation}
(\overline{\alpha},\overline{\beta},\overline{\sigma}_{Z})=\left(\alpha,\frac{\beta}{\gamma^{3+\alpha^{2}/4}},\sigma_{Z}+(2+\alpha)\log\gamma\right).\label{eq:scaling-Zparam}\end{equation}
Thus demanding $\overline{\sigma}_{Z}=0$ rescales $\beta$ to $\overline{\beta}=e^{-\frac{\sigma_{Z}}{1+\alpha}\left(3+\frac{\alpha^{2}}{4}\right)}\beta$. 

\begin{figure}
\includegraphics[width=0.5\columnwidth]{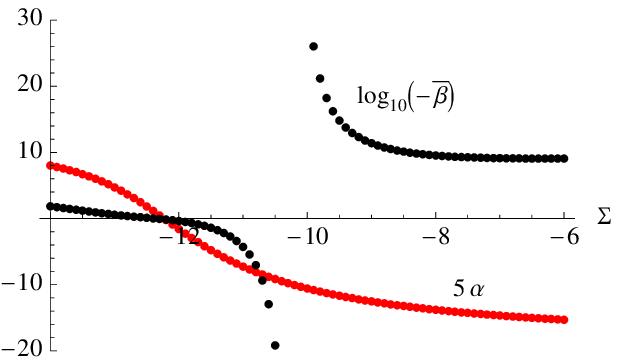}\includegraphics[width=0.5\columnwidth]{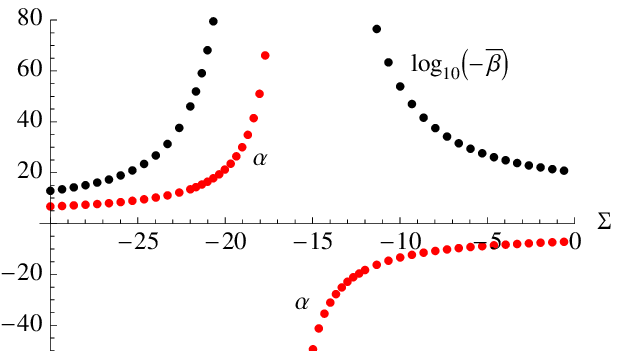}\caption{Parameters of the solution near the singularity \eqref{eq:Z-sol}
connecting to asymptotic solution \eqref{eq:asymp-sol} with $\sigma_{0}=0$.
These follow the two blue lines on Figure \ref{fig:parameters-at-infinity}.
On the left, $\mu_{0}=-10$, where there is never a horizon, and on
the right, $\mu_{0}=5$, crossing the curve of solutions with a regular
horizon at $\Sigma\approx-17$. The scaled parameter $\bar{\beta}$
is large and negative, so we plot $\log_{10}(-\bar{\beta})$, and
$5\alpha$ or $\alpha$ (in red). \protect \\
Note that, on the left, the point at which $\bar{\beta}$ diverges
is $\alpha=-2$. This is $c_{1}=\alpha+2=0$, in terms of the parameter
used in the Appendix. \label{fig:parameters-zero-from-A}}

\end{figure}

We can also start far away, with the asymptotic solution, and integrate
in to the singularity. For each point on $(\mu_{0},\Sigma)$ plane
of Figure \ref{fig:parameters-at-infinity} not on the curve drawn
there, we do not encounter a horizon before reaching the singularity.
Figure \ref{fig:parameters-zero-from-A} shows the resulting parameters
near the singularity, taking points $(\mu_{0},\Sigma)$ along the
two vertical lines drawn on Figure \ref{fig:parameters-at-infinity}.
These show the generic behaviour, either crossing the curve of solutions
with a regular horizon (at $\mu_{0}=5$) or avoiding it (at $\mu_{0}=-10$).

\section{Conclusions}

At spatial infinity the solution appears to have three parameters,
but since there is one scaling relationship, we can focus on two:
$\mu_{0}$ and $\Sigma$. Further restricting to solutions with a
regular horizon, we find a one-parameter family of solutions, with
horizon at $r=r_{H}\geq\sqrt{2}$. Only those with $r_{H}>2$ connect
smoothly to spatial infinity, which excludes the extremal case.

Numerically integrating outwards from the horizon, we find the curve
in the $(\mu_{0},\Sigma)$ plane of points which describe black holes.
Integrating inwards from the horizon, we arrive at a singularity at
$r=0$ without encountering another horizon. Starting at infinity,
from a point in the $(\mu_{0},\Sigma)$ plane not on the curve, we
can again integrate inwards all the way to the singularity at $r=0$,
showing that there isn't another class of horizons. 

Thus we concluded that the horizon structure of the 6-dimensional
black holes studied in \cite{Bergshoeff:2006bs} is like that of Schwarzschild;
unlike Reissner--Nordstr\"om and the six-dimensional Yang black holes
of \cite{Gibbons:2006wd} it does not have an inner horizon nor an
extremal limit.

\section*{Acknowledgments}

This research is supported in part by DOE grant DE-FG02-91ER40688-Task
A. 

\appendix

\section{Correction near the Singularity}

For the asymptotic solution \eqref{eq:asymp-sol} and the near-horizon
solution \eqref{eq:NH-sol}, it is clear that the next terms are higher
powers of $\frac{1}{r}$ or $d$, which are easy to calculate and
obviously smaller than the terms written. But for the solution \eqref{eq:Z-sol}
near $r=0$ things are not so simple. This appendix finds a correction
to this, in different variables, and shows that the remainder is small
in the appropriate limit.

We can combine the first two of equations \eqref{eq:eleven+twelve+thirteen}
into one third-order equation for $\sigma(r)$ alone by solving the
second for $\mu'(r)$ and substituting this into $\partial_{r}$ of
the first. The result is the quotient of a third-order and a second-order
differential equation \[
\frac{\mathcal{N}[\sigma'',\sigma',\sigma]}{\mathcal{D}[\sigma',\sigma]}=0\]
where the numerator and denominator are: \begin{align*}
\mathcal{N} & =r^{2}\left(e^{\sigma}r^{2}-2\right)\sigma'^{4}+r\left(\left(e^{\sigma}r^{2}-2\right)\sigma''r^{2}+12\right)\sigma'^{3}\\
 & \quad+4\left(4e^{\sigma}r^{2}+3\sigma''r^{2}-8\right)\sigma'^{2}-4r\left(e^{\sigma}r^{2}-2\right)\left(r\sigma'''-4\sigma''\right)\sigma'\\
 & \qquad-32\sigma''+8r\left(r\left(e^{\sigma}r^{2}-2\right)\sigma''^{2}-2\sigma'''\right)\,,\\
\mathcal{D} & =\frac{4}{3}e^{\sigma}\left(\sigma'+r\sigma''\right)\,.\end{align*}
In \eqref{eq:Z-sol} we had $\sigma(r)=\alpha\log r+\sigma_{Z}$,
which sets both $\mathcal{N}$ and $\mathcal{D}$ to zero. We will
refer to this as the singular solution.

We now seek a solution of $\mathcal{N}=0$ alone. This equation is
equivalent to the following nonlinear second-order equation: \begin{align}
0= & (-128x^{3}+32x^{4})b+(32x^{2}-8x^{3})b^{2}-(24x-6x^{2})b^{3}+(2-x)b^{4}\label{eq:second-order}\\
 & +\left[128x^{4}-32x^{5}+(32x^{2}-10x^{3})b^{2}-(2x-x^{2})b^{3}\right]b'\nonumber \\
 & +\left[-32x^{4}+8x^{5}-(8x^{3}-4x^{4})b\right]b'^{2}+\left[(-32x^{4}+8x^{5})b+(8x^{3}-4x^{4})b^{2}\right]b''\nonumber \end{align}
where the new function $b(x)$ is related to $\sigma(r)$ by\begin{equation}
e^{\sigma(r)}=\frac{x(r)}{r^{2}},\qquad r(x)=e^{\int\frac{1}{b(x)}dx+c_{3}}\,.\label{eq:sigma-to-b-rel}\end{equation}
One solution of this is $b_{0}(x)=c_{1}x$, leading to $\sigma(r)=(c_{1}-2)\log r-c_{1}c_{3}$,
the singular solution. To find a correction to this, we make the ansatz
\begin{equation}
b(x)=c_{1}x\left(1+\int_{x_{0}}^{x}p(x')dx'\right)\label{eq:b-ansatz}\end{equation}
and expand in $p$. Write the second order equation \eqref{eq:second-order}
as \[
0=\mathcal{A}[p(x)]+\mathcal{B}[p(x)]\]
 where $\mathcal{A}$ is the term linear in $p$, and $\mathcal{B}$
is the nonlinear piece. The linear equation $\mathcal{A}[p(x)]=0$
is \begin{align*}
0 & =c_{1}(c_{1}-4)x^{5}\left((8-6c_{1})(x-4)+c_{1}^{2}(x-2)\right)p(x)\\
 & \quad+4c_{1}^{2}x^{6}\left(8-2x+c_{1}(x-2)\right)p'(x)\end{align*}
which can be integrated to give\[
p_{1}(x)=c_{2}\, x^{\frac{c_{1}}{4}+\frac{4}{c_{1}}}\left(\frac{1}{x^{2}}(c_{1}-2)-\frac{2}{x^{3}}(c_{1}-4)\right)\,.\]
When substituting back into the ansatz \eqref{eq:b-ansatz} for $b(x)$,
the coefficient of $x$, initially $c_{1}$, will in general be renormalized
by the term from the lower limit of integration. Keeping this coefficient
free of $c_{2}$ will keep the power $x^{\frac{c_{1}}{4}+\frac{4}{c_{1}}}$
independent of $c_{2}$, allowing for a simple $c_{2}\to0$ limit.
To do this, we take \[
x_{0}=\frac{2c_{1}^{2}-8c_{1}+32}{(c_{1}-4)(c_{1}-2)}\]
and then obtain \begin{align}
b(x) & =c_{1}x+c_{2}x^{\frac{c_{1}}{4}+\frac{4}{c_{1}}}\left(\frac{4c_{1}^{2}(c_{1}-2)}{16+c_{1}(c_{1}-4)}+\frac{1}{x}\,\frac{8c_{1}^{2}}{4-c_{1}}\right)\,.\label{eq:c123-sol}\end{align}
This is, implicitly, an approximate solution for $\sigma(r)$ with
three parameters $c_{1}$, $c_{2}$ and $c_{3}$, thus cannot be a
solution to $\mathcal{D}=0$ for generic values. We can match it to
numerical $\sigma(r)$ using \eqref{eq:sigma-to-b-rel}.

The variable $x$ can be either large or small near $r=0$, depending
on the sign of $c_{1}$. From \eqref{eq:sigma-to-b-rel}, we have
$x(r)=r^{2}e^{\sigma(r)}\sim r^{2}r^{\alpha}=r^{c_{1}}$. When $c_{1}>0$,
$x\to0$ as $r\to0$. This occurs behind the regular horizon (see
Figure \ref{fig:parameters-at-zero}) and also for the naked singularity,
when $\Sigma\lesssim-10$ and $\Sigma\lesssim-10$ in Figure \ref{fig:parameters-zero-from-A}'s
first and second graphs. In the solution $b_{1}(x)$ above, the power
$\frac{c_{1}}{4}+\frac{4}{c_{1}}$ is at least 2, thus the $c_{2}$
correction terms die faster than the unperturbed term $c_{1}x$. 

When $c_{1}<0$, then instead $x\to\infty$ as $r\to0$. This occurs
for the other half of \ref{fig:parameters-zero-from-A}a. The power
$\frac{c_{1}}{4}+\frac{4}{c_{1}}$ is now negative, so the $c_{2}$
correction terms are small at large $x$, while the unperturbed term
is large. 

Finally, we perform a check that this approximation $p_{1}(x)$ to
the nonlinear solution is a good one. Define a Green's function $\mathcal{A}[G(x,y)]=\delta(x-y)$.
Then the full solution is \[
p(x)=p_{1}(x)-\int dy\, G(x,y)\,\mathcal{B}[p(y)]\,.\]
For $x<1$, the Green's function is\begin{align*}
G(x,y) & =\frac{8-2x+c_{1}(x-2)}{4c_{1}^{2}\left(8-2y+c_{1}(y-2)\right)^{2}x^{3}y^{3}}\left(\frac{x}{y}\right)^{\frac{c_{1}}{4}+\frac{4}{c_{1}}}\quad\mbox{when }x<y<1\mbox{, else zero}\\
 & \sim x^{-3}x^{\frac{c_{1}}{4}+\frac{4}{c_{1}}}\: y^{-3}y^{-(\frac{c_{1}}{4}+\frac{4}{c_{1}})}\quad\mbox{as }x,y\to0\end{align*}
and we approximate the full nonlinear kernel $\mathcal{B}[p(y)]$
by the known $\mathcal{B}[p_{1}(y)]$ to obtain the next order correction.
We find \[
\mathcal{B}[p_{1}(y)]\sim\left(y^{\frac{c_{1}}{4}+\frac{4}{c_{1}}}\right)^{2}\quad\mbox{as }y\to0\,.\]
Thus the nonlinear corrected solution looks like \begin{align*}
p(x) & \sim p_{1}(x)-x^{-3}x^{\frac{c_{1}}{4}+\frac{4}{c_{1}}}\int_{x}^{1}dy\, y^{-3}y^{\frac{c_{1}}{4}+\frac{4}{c_{1}}}+\ldots\\
 & \sim x^{-3}x^{\frac{c_{1}}{4}+\frac{4}{c_{1}}}\left(1+x^{-2}x^{\frac{c_{1}}{4}+\frac{4}{c_{1}}}+\ldots\right)\,.\end{align*}
Since $\frac{c_{1}}{4}+\frac{4}{c_{1}}>2$ (for $c_{1}\neq4$) this
is a small correction to the linear piece in the limit $x\to0$.

For $x>1$, the Green's function is\begin{align*}
G(x,y) & =-\frac{8-2x+c_{1}(x-2)}{4c_{1}^{2}\left(8-2y+c_{1}(y-2)\right)^{2}x^{3}y^{3}}\left(\frac{x}{y}\right)^{\frac{c_{1}}{4}+\frac{4}{c_{1}}}\quad\mbox{when }1<y<x\mbox{, else zero}\\
 & \sim x^{-2}x^{\frac{c_{1}}{4}+\frac{4}{c_{1}}}\: y^{-4}y^{-(\frac{c_{1}}{4}+\frac{4}{c_{1}})}\quad\mbox{as }x,y\to\infty\end{align*}
and the leading power in $\mathcal{B}$ is now \[
\mathcal{B}[p_{1}(y)]\sim y^{3}\left(y^{\frac{c_{1}}{4}+\frac{4}{c_{1}}}\right)^{2}\quad\mbox{as }y\to\infty\,.\]
This time the corrected solution is \begin{align*}
p(x) & \sim p_{1}(x)-x^{-2}x^{\frac{c_{1}}{4}+\frac{4}{c_{1}}}\int_{1}^{x}dy\, y^{-2}y^{\frac{c_{1}}{4}+\frac{4}{c_{1}}}+\ldots\\
 & \sim x^{-2}x^{\frac{c_{1}}{4}+\frac{4}{c_{1}}}\left(1+x^{-1}x^{\frac{c_{1}}{4}+\frac{4}{c_{1}}}+\ldots\right)\end{align*}
The nonlinear piece is now a very small correction in the limit $x\to\infty$,
as $\frac{c_{1}}{4}+\frac{4}{c_{1}}<-2$.

\bibliographystyle{utphys}
\bibliography{monopoles}

\end{document}